\documentclass[usenatbib,useAMS,twocolumn]{mn2e}
\usepackage{graphicx}
\usepackage{bm}
\usepackage{amsmath}
\usepackage[dvipdfm]{color}
\input{colordvi.tex}
\usepackage{color}
\usepackage[dvipdfm,
 bookmarkstype=toc=true,
 linktocpage=true,
 bookmarks=true
 ]{hyperref}
\usepackage[varg]{txfonts}
\def\bs#1{\mbox{\boldmath $#1$}}
\def\ave#1{\langle #1 \rangle}

\newcommand{\simgt}{\lower.5ex\hbox{$\; \buildrel > \over \sim \;$}}
\newcommand{\simlt}{\lower.5ex\hbox{$\; \buildrel < \over \sim \;$}}

\newcommand{\id}{{\rm d}}
\def\spc#1{\hspace{0.3cm} #1 \hspace{0.0cm}}

\begin{document}

\title[Nonlinear halo power spectrum]
      {Perturbation theory for nonlinear halo power spectrum: the
      renormalized bias and halo bias}

\author[A. J. Nishizawa et al.]
{Atsushi J. Nishizawa\thanks{email: atsushi.nishizawa@ipmu.jp}, 
Masahiro Takada 
and Takahiro Nishimichi\\
Kavli Institute for the Physics and Mathematics of the Universe 
(Kavli IPMU, WPI), 
The University of Tokyo, 
Chiba 277-8582, Japan}

\date{\today}

\maketitle

\begin{abstract}
  We revisit an analytical model to describe the halo-matter
  cross-power spectrum and the halo auto-power spectrum in the weakly
  nonlinear regime, by combining the perturbation theory (PT) for
  matter clustering, the local bias model, and the halo bias.
  Nonlinearities in the power spectra arise from the nonlinear
  clustering of matter as well as the nonlinear relation between the
  matter and halo density fields. By using the ``renormalization''
  approach, we express the nonlinear power spectra by a sum of the two
  contributions: the nonlinear matter power spectrum with the
  effective linear bias parameter, and the higher-order PT spectra
  having the halo bias parameters as the coefficients.  The halo
  auto-power spectrum includes the residual shot noise contamination
  that needs to be treated as additional free parameter.  The
  higher-order PT spectra and the residual shot noise cause a
  scale-dependent bias function relative to the nonlinear matter power
  spectrum in the weakly nonlinear regime.  We show that the model
  predictions are in good agreement with the spectra measured from a
  suit of high-resolution $N$-body simulations up to $k\simeq
  0.2~h{\rm Mpc}^{-1}$ at $z=0.35$, for different halo mass bins.
\end{abstract}

\begin{keywords} 
galaxies: clusters: general -- cosmology: theory -- dark energy -- large-scale structure of Universe
\end{keywords}

%=====================================================================================
%=====================================================================================
\section{Introduction}
\label{sec:intro}
%=====================================================================================
%=====================================================================================

Clustering statistics of galaxies such as the two-point correlation
function of galaxies or the Fourier-transformed-counterpart power
spectrum are powerful tools to constrain cosmology. In particular, the
baryon acoustic oscillation (BAO) experiment with 
wide-area galaxy redshift survey is recognized as a robust probe of
cosmological distances. There are various on-going and planned galaxy
surveys aimed at achieving high-precision BAO measurements over a
wider range of redshifts: the SDSS-III Baryon Oscillation
Spectroscopic Survey
(BOSS)\footnote{http://www.sdss3.org/surveys/boss.php}, Subaru Prime
Focus Spectrograph (PFS)
Survey\footnote{http://sumire.ipmu.jp/en/2652}, and the ESA Euclid
satellite
experiment\footnote{http://sci.esa.int/science-e/www/area/index.cfm?fareaid=102}.

The BAO scale is one particular length scale measured from the pattern
of galaxy clustering.  Much more significant signal-to-noise ratios
are inherent in the broad-band shape and amplitude information of the
galaxy power spectrum at BAO scales.  However, to reliably use the
amplitude information, we need to resolve various systematic
uncertainties in the weakly nonlinear regime: the nonlinear clustering
effect and galaxy bias uncertainty. There are promising developments
towards a more accurate modeling of the nonlinear clustering of matter
based on $N$-body simulations
\citep{Springeletal:05,Anguloetal:08,Takahashietal:09,Nishimichi:2009,Valageas:2011}
as well as perturbation theory (PT) of structure formation
\citep{Juszkiewicz:81,Vishniac:83,Makinoetal:92,JainBertschinger:94,JeongKomatsu:06,
CrocceScoccimarro:06,Matsubara:08,TaruyaHiramatsu:08,Saitoetal:08,Nishimichi:2007}.

The galaxy bias uncertainty is a harder problem, because physical
processes involved in galaxy formation/evolution are highly nonlinear
and still very challenging to model from the first principles. Hence a
practical approach often used assumes an empirical parametrization of
galaxy bias; the peak bias model
\citep{Kaiser:84,MoWhite:96,ShethTormen:99,Schmidtetal:12} and the
local bias model assuming a ``local'' mapping relation between galaxy
and matter distributions at each spatial position
\citep{Coles:93,FryGaztanaga:93,ScherrerWeinberg:98,Schmidtetal:12}.
In the peak bias model, a galaxy or more precisely halo is assumed to
form at or around the peak of the initial matter density field, where
a typical scale of the peaks corresponds to scales of the halo that
host galaxies at low redshifts (although one halo can contain several
galaxies inside). Thus the distribution of halos are by nature biased
relative to the underlying matter distribution \citep{Kaiser:84},
because only the density peaks can be places to have halos today,
while the under-density regions or the density minima are very
difficult (or impossible) to form halos. Furthermore, the
long-wavelength perturbation mode in the initial density field causes
modulations in the heights of the small-scale peaks, and in turn alter
subsequent formation of halos at low redshifts.  The amount of halo
bias can depend on the long-wavelength modes as a result of mode
coupling in the weakly nonlinear regime relevant for BAO scales.  The
halo bias or peak bias models have been studied in the literature
\citep{MoWhite:96,ShethTormen:99,Desjacquesetal:11,Scoccimarroetal:12,Schmidtetal:12}.

The nonlinear effect on the galaxy power spectrum arises from two
effects: the nonlinear clustering of matter (mostly dark matter) and
the nonlinear bias relation between the galaxy and dark matter
distributions. At BAO scales in the weakly nonlinear regime, we expect
that the nonlinear clustering of galaxies can be accurately modeled by
incorporating the PT of structure formation, the local bias model
and/or the peak-background split bias model (halo bias). Such an
attempt was first made in \citep{Heavens:1998}, and followed by
various works
\citep{Mcdonald:2006,Matsubara:08,JeongKomatsu:09,Saitoetal:09,McDonaldRoy:09,
  Maneraetal:10,baldaufetal:2010,Pollacketal:12,SatoMatsubara:11,Chanetal:12,Baldaufetal:12},
which continuously show an improved understanding of the nonlinear
galaxy power spectrum.

In this paper, we revisit the method of modeling the nonlinear power
spectra of halos, more explicitly halo-matter and halo-halo power
spectra, by incorporating the PT, the local bias model and the halo
bias. In doing this, we employ ``renormalization approach'' developed
in \citet{Mcdonald:2006} in order to re-sum contributions of the
nonlinear matter clustering up to the higher-order loop
corrections. This yields the term expressed by the product of the
``full'' nonlinear matter power spectrum and the renormalized linear
bias parameter. The remaining terms, for which we keep the one-loop
correction order based on the standard PT, give the effect of
scale-dependent halo bias in the halo power spectrum.  Thus our
approach fully utilizes the recent improvement in modelling the
nonlinear matter power spectrum (in this paper we will use the refined
PT prediction developed in \citet{Taruyaetal:09}).  We test the
accuracy of our model predictions by comparing with the halo spectra
measured from high-resolution $N$-body simulations done in
\citet{NishimichiTaruya:2011}.

This paper is organized as follows. In Sec.~\ref{sec:modeling}, we
develop a method of modeling the nonlinear halo-matter and halo-halo
power spectra by incorporating the PT, the local bias model and the
halo bias. In this section, we also show the detailed comparison of
the model predictions with the simulation results for the halo
catalogs of different mass bins.  Throughout this paper we assume the
$\Lambda$ dominated, cold dark matter model ($\Lambda$CDM) as for our
fiducial model which is consistent with \citet{Komatsu.etal:2011}: the
density parameters of matter, baryon and the cosmological constant are
$\Omega_{\rm m0}=0.279$, $\Omega_{\rm b0}/\Omega_{\rm m0}=0.165$, and
$\Omega_{\Lambda}=0.721$ (i.e. flat geometry), the Hubble parameter
$h=0.701$, the tilt of the primordial power spectrum $n_s=0.96$ and
the power spectrum normalization $\sigma_8=0.817$.

%=====================================================================================
%=====================================================================================
\section{Preliminaries: perturbation theory and halo bias}
\label{sec:modeling}
%=====================================================================================
%=====================================================================================

Our model is based on three ingredients; the perturbation theory (PT)
of structure formation
\citep{Juszkiewicz:81,Vishniac:83,Fry:84,Goroffetal:86,SutoSasaki:91,JainBertschinger:94},
the local bias model
\citep{Coles:93,FryGaztanaga:93,ScherrerWeinberg:98} and the halo bias
model \citep{MoWhite:96,ShethTormen:99} \citep[also see][for thorough
reviews]{CooraySheth:02,Bernardeauetal:02}.  In this section, we
briefly review the PT and the halo bias we will employ in the
following sections.

%=====================================================================================
\subsection{Standard Perturbation Theory}
%=====================================================================================

Throughout this paper, we consider a pressure-less, irrotational fluid
system and assume cold dark matter as the dominant fluid component to
drive gravitational instability of structure formation. The nonlinear
dynamics in an expanding universe is fully characterized by the
density fluctuation field, $\delta_m$, and the peculiar velocity field
$\theta_m$ \citep{Bernardeauetal:02}.  Given the initial conditions,
the time evolutions of the fields are governed by the continuity
equation, the equation of motion and the Poisson equation.

By using the standard PT, we can solve the nonlinear dynamics.  The
density fluctuation field at a given redshift $z$ is expanded as
\begin{eqnarray}
  \label{eq:SPT}
  \delta_m(\bs{k},z)
&=&\delta_{m(1)}(\bs{k},z)+\delta_{m(2)}(\bs{k},z)
+\delta_{m(3)}(\bs{k},z)+\cdots.
\end{eqnarray}
The PT solution for the $n$-th order density fluctuation field can be
found to be
\begin{eqnarray}
\delta_{m(n)}(\bs{k},z)  &\equiv &
   D_{+}^n(z)
  \int \id^3\bs{q}_1 \cdots \id^3\bs{q}_n
  \delta_{m(1)}(\bs{q}_1)\cdots\delta_{m(1)}(\bs{q}_n) \nonumber \\
  && \hspace{1.0cm}
  \times F_n(\bs{q}_1,\cdots,\bs{q}_n)
  \delta_D^3\left(\bs{k}-\sum_i\bs{q}_i\right),
\end{eqnarray}
where $\delta_{m(1)}$ is the linear density field today, $D_+(z)$ is
the linear growth rate normalized as $D_+(z=0)=1$, and
$\delta_D^3(\bs{k})$ is the Dirac delta function.  The $n$-th order
density fluctuation field has the amplitude of the order
${\mathcal O}[(\delta_{m(1)})^n]$.  The growth rate can be computed, e.g. by
solving Eq.~(7) in \citet{OguriTakada:2011}.  The Fourier kernel
$F_n(\bs{q}_1,\cdots,\bs{q}_n)$ describes a coupling between different
Fourier modes due to nonlinear clustering, and we will use the
expression in Eq.~(10a) of \citet{JainBertschinger:94}. Note that,
although the form of the Fourier kernel is exact only for an Einstein
de-Sitter model with $\Omega_{m}=1$, it was shown to be a good
approximation of the exact solution for a $\Lambda$CDM model.

%=====================================================================================
\subsection{Halo Mass Function and Halo Bias}
%=====================================================================================
\begin{figure}
  \includegraphics[width=\linewidth]{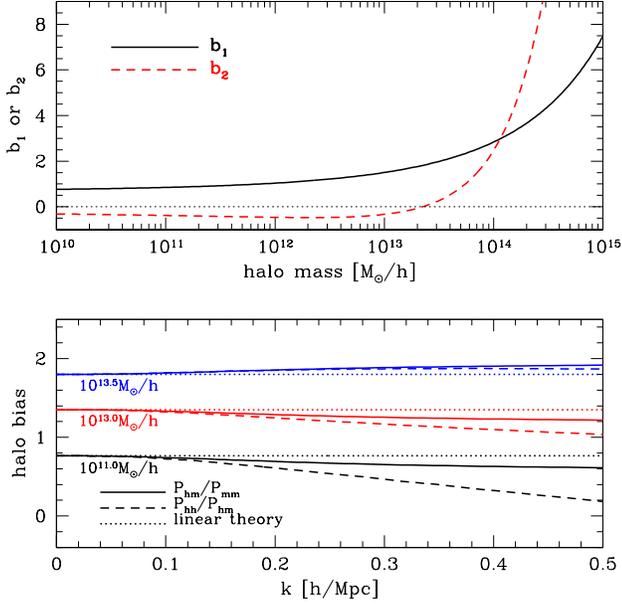}
  \caption{
    {\em Upper panel}:  
    the halo bias parameters, $b_1(M)$ and $b_2(M)$, as a function of halo
    mass, computed from Eq.~(\ref{eq:bias_theory}). We consider 
    $\Lambda$CDM model and redshift $z=0.35$. The nonlinear bias parameter,  
    $b_2$,  is negative for low mass halos with $M \simlt 2.5\times 10^{13}$, 
    while it becomes positive and rapidly increases for the more massive halos.
    {\em Lower panel}: 
    the renormalized PT prediction for halo bias functions for halos with 
    masses $M=10^{11}, 10^{13}$ and $10^{13.5} M_{\odot}/h$ (from bottom to
    top curves), respectively.  We used Eqs.~(\ref{eq:phm_renorm}) and
    (\ref{eq:phh_renorm}) to compute these curves. Note that
    we set the effective bias parameter to $b_1^{\rm eff}=0.9b_1(M)$ 
    (0.9 times the linear halo bias parameter) as implied
    from the simulations (see Sec.~\ref{sec:results}), and set the
    residual shot noise term to $\delta N=0$ for simplicity.
    \label{fig:bias}
  }
\end{figure}

Dark matter halos that host galaxies and/or galaxy clusters are useful
tracers of large-scale structure, and can be used to infer the
underlying dark matter distribution. However, the halo and dark matter
distributions are not the same, which leaves an uncertainty, the
so-called bias uncertainty. In this paper we employ the halo bias
model developed in \citet{MoWhite:96,ShethTormen:99} \citep[also
see][]{CooraySheth:02}.

Let us start with defining the halo mass function $n(M,z)dM$, which
gives the comoving number density of halos in the mass range
$[M,M+dM]$ and at redshift $z$.  We employ the mass function given in
\citet{ShethTormen:99}:
\begin{eqnarray}
n(M)\id M&=&\frac{\bar{\rho}_{m0}}{M}f(\nu)\id\nu\nonumber\\
&=&\frac{\bar{\rho}_{m0}}{M}A\left[1+(a\nu)^{-p}\right]
\sqrt{a\nu}\exp\left[-\frac{a\nu}{2}\right]\frac{d\nu}{\nu}, 
\label{eq:dndm}
\end{eqnarray}
where $\bar{\rho}_{m0}$ is the mean mass density today; $\nu\equiv
[\delta_c/D_+(z)\sigma_m(M)]^2$; $\delta_c$ is the threshold
over-density for spherical collapse model; $\sigma_m(M)$ is the
present-day rms fluctuations in the mass density top-hat smoothed over
scale $R=(3M/4\pi\bar{\rho}_{m0})^{1/3}$. We will throughout this
paper employ the coefficients $a=0.75$ and $p=0.3$, which are obtained
by comparing the fitting formula to $N$-body simulations. Note that
the normalization coefficient $A$ is determined so as to satisfy the
normalization condition $\int_0^\infty\! \id\nu f(\nu)=1$.

The mass function above holds only in an ensemble average sense,
i.e. the average of the halo distribution over a sufficiently large
volume. In other words, the number density of halos in a finite volume
is modulated according to fluctuations of the underlying matter
distribution within the volume, $\delta_m$.  Employing the local bias
model for halos, we assume that the halo distribution at a given
spatial position $\bs{x}$ is locally related to the underlying matter
distribution at the position $\bs{x}$ as
\begin{eqnarray}
\delta_{\rm h}(\bs{x},z)&=&F(\delta_{m}(\bs{x},z))\nonumber\\
&=&\sum_{n=0}^\infty\frac{1}{n!}F^{(n)}(\delta_m=0)
\left\{[\delta_m(\bs{x},z)]^n
-\left\langle[\delta_m(\bs{x},z)]^n\right\rangle
\right\},
\label{eq:local_bias_def}
\end{eqnarray}
where $F$ is the functional to govern the local mapping relation. In
the second line of the r.h.s., we have Taylor-expanded the relation in
terms of $\delta_m(\bs{x})$, and $F^{(n)}$ denotes the $n$-th
derivatives of $F$ with respect to $\delta_m$.  Exactly speaking, as
stressed in \cite{Schmidtetal:12}, the local bias relation would hold
to a good approximation in a ``peak-background split'' picture
\citep[also see][for the pioneer work]{MoWhite:96,ShethTormen:99}. In
the peak-background model, the matter density field is divided into
long- and short-wavelength modes, which correspond to ``background''
and ``peak'' density modes, respectively. The short-wavelength modes
are at the scales responsible for formation of halos corresponding to
about 10Mpc at maximum in the initial density fields and therefore are
well below BAO scales (up to $k\sim $ a few 0.1~$h/{\rm Mpc}^{-1}$ in
wavenumber). The long-wavelength modes are a ``coarse-grained'' field
responsible for a modulation of the peak heights of short-wavelengths,
and in this paper we assume that the long-wavelength modes are at BAO
scales.  Hence we assume that $\delta_m(\bs{x})$ in
Eq.~(\ref{eq:local_bias_def}) is the coarse-grained field, even though
we did not explicitly denote a notation to express the smoothing
nature of $\delta_m(\bs{x})$.  The term
$\left\langle[\delta_m(\bs{x})]^n\right\rangle$ in the above equation
is introduced to enforce $\langle\delta_{\rm h}\rangle=0$.

As shown in \cite{Schmidtetal:12}, the expansion coefficients in
Eq.~(\ref{eq:local_bias_def}), $F^{(n)}$, can be related to the
peak-background split bias parameters or halo bias parameters in an
ensemble average sense.  Since we focus on the halo correlation
functions in this paper, we {\em empirically} assume that the halo
density field in Fourier space is given as
\begin{eqnarray}
  \label{eq:halo_expansion}
  \delta_{\rm h}(\bs{k},z)
  &=&\sum_n \frac{b_n}{n!}\int\! \id^3\bs{q}_1\cdots \id^3\bs{q}_n
  \delta_D^3\left(\bs{k}-\sum_i \bs{q}_i\right) \nonumber \\
  && \hspace{1em}
  \times \delta_{m}(\bs{q}_1,z)\cdots
  \delta_{m}(\bs{q}_n,z)+\epsilon(\bs{k}),
\end{eqnarray}
where $b_n$ is the halo bias parameters and we have set $F^{(n)}=b_n$
when converting Eq.~(\ref{eq:local_bias_def}) to the above equation.
The 1st and 2nd-order bias parameters, which are relevant for the
results in the following sections, are given in terms of the
derivatives of halo mass function (Eq.~\ref{eq:dndm}):
\begin{eqnarray}
  b_1(M)&=&1+c_1+E_1\nonumber\\
  b_2(M)&=&2\left(1-\frac{17}{21}\right)(c_1+E_1)+c_2+E_2,
  \label{eq:bias_theory}
\end{eqnarray}
where 
\begin{eqnarray}
&&c_1= \frac{a\nu-1}{\delta_c},\hspace{1em}c_2=\frac{a\nu}{\delta_c^2}
\left(a\nu-3\right), \nonumber\\
&&E_1=\frac{2p/\delta_c}{1+(a\nu)^p},\hspace{2em}
\frac{E_2}{E_1}=\frac{1+2p}{\delta_c}+2c_1.
\end{eqnarray}
In Eq.~(\ref{eq:halo_expansion}), to keep more generality, we included
the additional term $\epsilon(\bs{k})$ to model the noise field that
is uncorrelated to the matter density field,
i.e. $\langle\epsilon\delta_m\rangle=0$ \citep[see][]{Mcdonald:2006}.
The term $\left\langle[\delta_m(\bs{x})]^n\right\rangle$ in
Eq.~(\ref{eq:local_bias_def}) contributes only to the monopole mode of
$k=0$, so we ignored the contribution as it is not relevant for the
halo power spectra.

We again notice that Eq.~(\ref{eq:halo_expansion}) is not exact, and
rather ansatz we employ in this paper. We will test how well our
empirical, analytical model can describe the halo power spectra in the
weakly nonlinear regime by comparing the model predictions with the
simulation results.

%=====================================================================================
%=====================================================================================
\section{Renormalized perturbation theory for nonlinear halo power spectra}
\label{sec:renorm}
%=====================================================================================
%=====================================================================================

In this section, we model nonlinear cross-power spectrum of matter and
halos and nonlinear auto-power spectrum of halos by combining the
``renormalized'' PT approach
\citep{Mcdonald:2006,Saitoetal:09,JeongKomatsu:09,Saitoetal:11} with
the local bias model, the halo bias and the perturbation theory
described in the preceding section.

%=====================================================================================
\subsection{Halo-matter cross-power spectra}
\label{ssec:hm_auto_power}
%=====================================================================================
First, let's consider the matter power spectrum defined as 
\begin{equation}
\ave{\delta_m(\bs{k})\delta_m(\bs{k}')}\equiv (2\pi)^3
P_m(k)\delta_D^3(\bs{k}+\bs{k}'). 
\end{equation}
Using Eq.~(\ref{eq:SPT}), we can find that the power spectrum
including up to the one-loop corrections are given as
\begin{equation}
P_m(k;z)=P_m^{\rm L}(k;z) + P_{m(13)}(k;z)+P_{m(22)}(k,z), 
\end{equation}
where $P_m^{\rm L}(k;z)$ is the linear power spectrum, and $P_{m(13)}$
and $P_{m(22)}$ are the one-loop corrections arising from the ensemble
averages of the higher-order matter density fluctuation fields;
$\langle\delta_{m(1)}\delta_{m(3)}\rangle$ and
$\langle\delta_{m(2)}\delta_{m(2)}\rangle$, respectively. The one-loop
corrections at a given redshift $z$ can be computed once the linear
power spectrum at the redshift is specified:
\begin{eqnarray}
P_{m(13)}
&\equiv& 
  \frac{k^3P_m^{\rm L}(k;z)}{252(2\pi)^2}
  \int_0^{\infty}\!\id rP_m^{\rm L}(kr;z)\left[
  \frac{12}{r^2}-158+100r^2\right.\nonumber\\
&&\left.-42r^4+\frac{3}{r^2}(r^2-1)^3(7r^2+2)
  \ln\left|\frac{1+r}{1-r}\right|
  \right],\nonumber\\
P_{m(22)}
&\equiv& 
  \frac{k^3}{98(2\pi)^2}
  \int_0^{\infty}\!\id rP_m^{\rm L}(kr;z)\nonumber\\
&&\hspace{-3em}\times
  \int^1_{-1}\!\id\mu P_m^{\rm L}(k\sqrt{1+r^2-2r\mu};z)
  \frac{(3r+7\mu-10r\mu^2)^2}{(1+r^2-2r\mu)^2}.
\end{eqnarray}

Similarly, using the standard PT and halo bias prescription, we can
compute the cross-power spectrum between matter and halos of mass $M$,
which is the quantity that halo-shear cross-correlation can directly
probe. By inserting Eq.~(\ref{eq:SPT}) into
Eq.~(\ref{eq:halo_expansion}), we can find the formal expression of
the halo-matter cross-spectrum in a self-consistent manner by
including up to the one-loop correction terms of ${\mathcal O}(\delta_{m(1)}^4)$:
\begin{eqnarray}
  \label{eq:pgh_pt}
P_{{\rm h}m}(k;M,z ) &=&
  \left[ 
    b_1 +
    \frac{\sigma^2}{2}\left(
      b_3 + \frac{68}{21}b_2 
    \right) 
  \right] 
  P^{\rm L}_m(k) \nonumber\\
  &&\hspace{-3em}+ b_1\left[P_{m(13)}(k)+P_{m(22)}(k)\right] 
\nonumber\\
  &&\hspace{-3em}
+ {b}_2 \int
    \frac{\id^3\bs{q}}{(2\pi)^3}\! P^{\rm L}_m(q)P^{\rm L}_m(|\bs{k}-\bs{q}|)
    F_{2}(\bs{q},\bs{k}-\bs{q}),
\end{eqnarray}
where $\sigma^2\equiv \int\! \id^3\bs{q}/(2\pi)^3 P^{\rm L}_m(q)$ and we
employed notational simplification in the halo bias parameters;
$b_i=b_i(M)$. Thus, a formal implementation of the standard perturbation
theory (SPT)-based
halo-matter spectrum yields the divergence term, i.e $\sigma^2\sim
\int^\infty \!  q^3 P^{\rm L}_m(q)\id\ln q \rightarrow \infty$, for a
CDM-type power spectrum. In practice, since halo formation involves a
coarse-grained smoothing of the underlying matter distribution
corresponding to halo scales (also see discussion below
Eq.~\ref{eq:local_bias_def}), the divergence does not arise in the
power spectrum we actually observe.  Also note that the prefactor
coefficient of the linear power spectrum $P_m^{\rm L}$ is independent
of $k$.

From the first two terms of the r.h.s. of Eq.~(\ref{eq:pgh_pt}), we
might re-write the two terms as
\begin{eqnarray}
&&  \left[ 
    b_1 +
    \frac{\sigma^2}{2}\left(
      b_3 + \frac{68}{21}b_2 
    \right) 
  \right] 
  P^{\rm L}_m(k) 
+ b_1\left[P_{m(13)}(k)+P_{m(22)}(k)\right] \nonumber\\
&&\hspace{2em}=\left[b_1+\delta b_1\right] P^{\rm L}_m(k)
+b_1 \delta P_m(k) \nonumber\\
&&\hspace{2em}\simeq \left[b_1+\delta b_1\right]P_m^{\rm NL}(k) +
{\mathcal O}(\delta b_1 \delta P_{m}),
\label{eq:phm_renorm_step}
\end{eqnarray}
where we have defined the notations $\delta b_1\equiv
\sigma^2(b_3+68b_2/21)/2$ and $\delta P_m\equiv P_{m(13)}+P_{m(22)}$,
which are the higher-order contributions to the linear bias and the
linear matter power spectrum by the order of ${\mathcal O}(\delta_{m(1)}^2)$ 
with respect to leading order in the PT formalism.  
Motivated by the equation above as
well as the similar idea proposed by \citet{Mcdonald:2006}, we propose
the ``{\em renormalized}'' power spectrum as
\begin{eqnarray}
\label{eq:phm_renorm}
P_{{\rm h}m}(k;M, z)
&\equiv& 
b_1^{\rm eff}P^{\rm NL}_m(k; z)\nonumber\\
&&\hspace{-5em}+ {b}_2(M) \int
    \frac{\id^3\bs{q}}{(2\pi)^3}\! P^{\rm L}_m(q)P^{\rm L}_m(|\bs{k}-\bs{q}|)
    F_{2}(\bs{q},\bs{k}-\bs{q}).
\end{eqnarray}
The first term is given by the {\rm nonlinear} matter power spectrum,
$P_{m}^{\rm NL}$, multiplied by the ``effective'' or ``renormalized''
linear bias parameter, $b^{\rm eff}_1$, while the second term includes
the bare halo bias, $b_2(M)$, in Eq.~(\ref{eq:halo_expansion}).  Thus
the renormalized term can include the nonlinear corrections of matter
clustering up to the higher orders. From the PT viewpoint, this is not
self-consistent in a sense that the term $b_1^{\rm eff}P_m^{\rm NL}$
includes the higher-order contributions than the one-loop order.

There are several nice features in our renormalization prescription:
\begin{itemize}
\item At the limit of small $k$ or the linear regime, $P_{{\rm h}m}(k)
  \rightarrow b_1^{\rm eff}P^{\rm L}_m(k)$, because $P^{\rm NL}_{{\rm
      h}m}(k)\rightarrow P^{\rm L}_m(k)$ at the limit.
\item The effective linear bias $b_1^{\rm eff}$ is a parameter, and is
  not related to the linear halo bias $b_1(M)$ due to the
  renormalization.  Observationally, however, it can be determined by
  the cross-power spectrum at small $k$, e.g. measured from the
  large-scale signal of galaxy-galaxy weak lensing
  \citep{Mandelbaumetal:12}.
\item The scale-dependent bias, relative to the {\em nonlinear} mass
  power spectrum, arises from the term that depends on the
  second-order halo bias $b_2(M)$ and the linear power spectrum.
  Therefore the term is predictable once the background cosmological
  model, the halo masses and the redshift are specified.
\end{itemize}
We will test an accuracy of the renormalized cross-power spectrum
(Eq.~\ref{eq:phm_renorm}) by comparing the predictions with simulation
results for halos of various mass ranges.  We will below show how the
use of the ``full'' nonlinear matter power spectrum in
Eq.~(\ref{eq:phm_renorm}) can give a better fit to the simulation.

%=====================================================================================
\subsection{Halo auto-power spectrum}
\label{ssec:hh_auto_power}
%=====================================================================================
Similarly, we propose the {\em renormalized} auto-power spectrum of
halos with masses $M$ and $M'$ (for generality of discussion, we
consider the case that halos are in different mass ranges):
\begin{eqnarray}
\label{eq:phh_renorm}
P_{\rm hh'}(k; M, M',z)
  &=& 
    \left[ b_1b_1' + \frac{\sigma^2}{2} \left(b_1b_3^{\prime} + 
        \frac{68}{21}b_1 b_2^{\prime} \right)
\right.\nonumber\\
&&\hspace{-7em} +\left.
\frac{\sigma^2}{2} \left(b_1^{\prime }b_3 + 
\frac{68}{21}b_1^{\prime} b_2 \right)
\right]P^{\rm L}_m(k)
%\nonumber\\
% &&
+  b_1b_1^{\prime}\left[
P_{m(13)}(k)+P_{m(22)}(k)
\right]
 \nonumber \\
 & &\hspace{-7em}
    + \frac{1}{2}b_2b_2^{\prime}\int\! \frac{\id^3\bs{q}}
    {(2\pi)^3} P_m^{\rm L}(q)P_m^{\rm L}(|\bs{k}-\bs{q}|) \nonumber \\
  &&\hspace{-7em}
    + \left(b_1b_2^{\prime}+b_1^\prime b_2\right)
\int\!\frac{\id^3\bs{q}}{(2\pi)^3} P^{\rm L}_m(q)P^{\rm L}_m(|\bs{k}-\bs{q}|)
    F_2(\bs{q},\bs{k}-\bs{q}) 
%\nonumber\\
%&&\hspace{5em}
+ \delta N\nonumber\\
&&\hspace{-8em}\simeq b_1^{\rm eff}b_1^{{\rm eff}\prime}P^{\rm NL}_m(k)
\nonumber\\
&&\hspace{-7em}
+\frac{1}{2}b_2b_2^{\prime}\int\! \frac{\id^3\bs{q}}
    {(2\pi)^3} 
\left[
P_m^{\rm L}(q)P_m^{\rm L}(|\bs{k}-\bs{q}|) 
-P_m^{\rm L}(q)^2
\right]\nonumber \\
  &&\hspace{-7em}
    + \left(b_1b_2^{\prime}+b_1^\prime b_2\right)
\int\!\frac{\id^3\bs{q}}{(2\pi)^3} P^{\rm L}_m(q)P^{\rm L}_m(|\bs{k}-\bs{q}|)
    F_2(\bs{q},\bs{k}-\bs{q}) 
%\nonumber\\
%&&\hspace{5em}
+ \delta N',\nonumber\\
\end{eqnarray}
where we have introduced the effective linear bias $b_1^{\rm eff}$ and
$b_1^{{\rm eff}\prime}$ for halos of masses $M$ and $M'$,
respectively, and used the collapsed notations such as $b_2=b_2(M)$
and $b_2^\prime=b_2(M')$ and similarly those for $b_1$ and $b_1'$.
The last term $\delta N'$ denotes the residual shot noise term arising
from $\ave{\epsilon(\bs{k};M)\epsilon^*(\bs{k}';M')}$ in
Eq.~(\ref{eq:halo_expansion}) as follows.  As in the literature, we
refer to the term as the residual noise component after subtracting
the Poisson shot noise, $1/n_{\rm h}$, without addressing the origin.
Assuming that the residual shot noise is constant over the scale, but
may vary with halo masses, $\epsilon(\bs{k},M) \rightarrow
\epsilon(M)$, we will below study whether the model including the
residual shot noise can give a better fit to the $N$-body simulation
results for halos of different mass ranges.  In the second line on the
r.h.s. of the above equation, we re-defined the shot noise term as
$\delta N'=\delta N+b_2b_2'\int\!(\id^3\bs{q}/(2\pi)^3)P_{m}^{\rm
  L}(q)^2$ so that the following term (the second term in the second
line), which contributes to the scale-dependent bias, becomes finite
for the limit $k\rightarrow \infty$:
\begin{eqnarray}
&&\hspace{-1em}\frac{1}{2}b_2b_2^{\prime}\int\! \frac{\id^3\bs{q}}
    {(2\pi)^3} 
P_m^{\rm L}(q)P_m^{\rm L}(|\bs{k}-\bs{q}|)
\nonumber\\
&&\hspace{0em}\rightarrow 
\frac{1}{2}b_2b_2^{\prime}\int\! \frac{\id^3\bs{q}}
    {(2\pi)^3} 
\left[
P_m^{\rm L}(q)P_m^{\rm L}(|\bs{k}-\bs{q}|) 
-P_m^{\rm L}(q)^2
\right]. 
\end{eqnarray}
Since $\int\!\id^3\bs{q} P_m^{\rm L}(q)^2$ is constant, the residual
shot noise term modified in this way is still constant in space, but
can vary with halo masses both through the dependence on
$b_2(M)b_2(M')$ and $\epsilon$. The constant $\delta N'$ needs to be
treated as an additional free parameter for predicting $P_{\rm
  hh}(k)$.  In the following, we refer to the residual shot noise term
as $\delta N$, instead of $\delta N'$, for notational simplicity.  The
effective bias parameter $b_1^{\rm eff}$ in Eq.~(\ref{eq:phh_renorm})
is the same to that in Eq.~(\ref{eq:phm_renorm}) up to the order
${\mathcal O}(\delta_{m(1)}^2)$.

We would like to notice features of the
renormalized halo power spectrum:
\begin{itemize}
\item  At the limit of small $k$ or the linear regime, $P_{{\rm hh'}}(k)
\rightarrow b_1^{\rm eff}b_1^{\rm eff\prime}P^{\rm L}_m(k)+\delta N$.
\item The scale-dependent bias depends on the linear power spectrum
  and the halo bias parameters, $b_1(M)$ and $b_2(M)$, and therefore
  is predictable once the background cosmological model, the halo
  masses and the redshift are specified.
    \item The residual shot noise term $\delta N$ needs to be included and
      treated as a free parameter.
\end{itemize}
It is worth mentioning difference between our approach and
\cite{Mcdonald:2006}.  In \citet{Mcdonald:2006}, all the bias
coefficients are replaced with the renormalized bias parameters;
$b_1\rightarrow b_1^{\rm eff}$ and $b_2\rightarrow b_2^{\rm
  eff}$. Hence, the bias parameters need to be treated as free
parameters, and their relations with the halo bias parameters were not
discussed.  We will below test the validity of
Eq.~(\ref{eq:phh_renorm}) using simulations.

The lower panel of Fig.~\ref{fig:bias} shows the halo model
predictions for the effective bias functions, which are defined in
terms of the halo and matter power spectra: $b^{\rm cross}(k)\equiv
P_{{\rm h}m}(k)/P_{m}(k)$ or $b^{\rm auto}(k)\equiv P_{\rm
  hh}(k)/P_{{\rm h}m}(k)$ (see Eqs.~\ref{eq:phm_renorm} and
\ref{eq:phh_renorm}). Note that we did not include the residual shot
noise contribution in this plot (set $\delta N=0$).  Our model
predicts a scale dependent bias in the weakly nonlinear regime, and
the degree of scale-dependence changes with halo masses.  The top
panel of Fig.~\ref{fig:bias} shows that $b_2(M)$ is negative for low
mass halos, goes to zero around $M \simeq 2\times 10^{13} M_{\odot}/h$
and then becomes positive for more massive halos.  Hence,
Eqs.~(\ref{eq:phm_renorm}) and (\ref{eq:phh_renorm}) tell that the
nonlinear bias due to $b_2(M)$ suppresses the power spectrum
amplitudes for low mass halos, while it enhances for high mass halos.
Also the model shows that, unlike the linear theory prediction, the
scale-dependent halo bias generally differs in the estimators,
$P_{{\rm h}m}/P_m$ and $P_{\rm hh}/P_{{\rm h}m}$.  Hence $r\ne 1$ in
the weakly nonlinear regime, where $r$ is the correlation coefficient
of halo bias, defined as $r\equiv P_{\rm hh}/\sqrt{ P_{{\rm h}m}P_m}$.

%=====================================================================================
\subsection{HOD model: Relating halos to galaxies}
%=====================================================================================
Although we have focused on the halo power spectra, halos are not a
direct observable from a galaxy redshift survey and need to be
inferred from the distribution of galaxies. A practically useful
approach to relate galaxies to halos is using the halo occupation
distribution (HOD)
\citep{PeacockSmith:00,Seljak:00,Scoccimarroetal:01}.  In this paper,
for simplicity we assume that we can implement the method developed in
\citet{ReidSpergel:09} \citep[also
see][]{Reidetal:09,Reidetal:10,Hikageetal:12,Hikageetal:12b} for
reconstructing the halo distribution from the measured galaxy
distribution. In the halo catalog, each halo hosts only one galaxy.

In this setting, there is only one galaxy per halo.
The cross-power spectrum of matter and galaxies and the galaxy auto-power
spectrum are given in terms of the halo spectra as
\begin{eqnarray}
P_{gm}(k;z)&=&\frac{1}{\bar{n}_g}
\int\!\id M \frac{\id n}{\id M}N_{\rm HOD}(M) P_{{\rm h}m}(k; M,z),
\label{eq:pgm_renorm}
\\
P_{gg}(k;z)&=&\frac{1}{\bar{n}_g^2}
\int\!\id M 
\frac{\id n}{\id M}N_{\rm HOD}(M)
\nonumber\\
&&\hspace{-1em}\times
\int\!\id M'
\frac{\id n}{\id M'}N_{\rm HOD}(M')
 P_{{\rm h h'}}(k; M, M', z),
\label{eq:pgg_renorm}
\end{eqnarray}
where the spectra $P_{{\rm h}m}$ and $P_{\rm hh'}$ are given by
Eqs.~(\ref{eq:phm_renorm}) and (\ref{eq:phh_renorm}), respectively.
Here $N_{\rm HOD}(M)$ is the halo occupation distribution (HOD), and
$\bar{n}_g$ is the mean number density of galaxies, defined as
$\bar{n}_g\equiv \int\!\id M~ (\id n/\id M) N_{\rm HOD}(M)$.  Exactly speaking,
the equations above are valid only if each galaxy is at the center of
halo.  When the galaxies have an offset from the halo center, we need
to include a convolution of the average offset distribution of
galaxies, $\tilde{p}_{\rm off}(k; M)$, with the HOD distribution (see
\citet{Hikageetal:12,Hikageetal:12b} for details). However, for the
real-space power spectrum we focus on in this paper, the effect is
negligible on scales of interest; $\tilde{p}_{\rm off}(k; M)\approx 1$
at the scales of interest. For the redshift-space power spectrum, the
off-centered galaxies cause a significant Fingers-of-God effect. The
main focus of this paper is the scale-dependent galaxy bias due to
nonlinearities of structure formation, so we focus on the real-space
power spectrum.

Recently \citet{Hamausetal:11} proposed a method, more directly based
on the halo model approach, to model the nonlinear power spectrum of
halos. In this method, the nonlinear halo-matter and halo-halo power
spectra are given by a sum of the linear power spectrum, multiplied
with linear bias parameter, and the 1-halo term.  The 1-halo term of
$P_{hm}$ is calculated by the mass weighted integral of the mass
function, because the Fourier-transformed halo profile
$\tilde{\rho}(k,M)/M\simeq 1$ in the weakly nonlinear regime of
interest.  We will below compare the performance of our method and
theirs by comparing the model predictions with our own high-resolution
$N$-body simulations.

%=====================================================================================
%=====================================================================================
\section{Testing the nonlinear halo spectra with $N$-body simulations}
\label{sec:results}
%=====================================================================================
%=====================================================================================
\begin{table*}
  \begin{center}
  \begin{tabular}{l|rrrrrrrrr} \hline\hline
 Halo catalog & \spc{bin 1} & \spc{bin 2} & \spc{bin 3}  & \spc{bin 4} & \spc{bin 5}
              & \spc{bin 6} & \spc{bin 7} & \spc{bin 8}  & \spc{bin 9} \\ \hline
   $M_{\rm min}/10^{12}~[M_\odot/h]$ 
%%%%  FOF Mass 
%             & 1.77 & 2.49 & 3.54 & 4.98 & 7.09 
%             & 10.0 & 14.2 & 20.1 & 28.4 \\
%%%% scaled virial Mass
             & 1.55 & 2.18 & 3.11 & 4.37 & 6.22 
             & 8.77 & 12.5 & 17.6 & 24.9 \\
   $M_{\rm max}/10^{12}~[M_\odot/h]$  
%%%%  FOF Mass 
%             & 5.54 & 10.2 & 17.4 & 26.6 & 40.4 
%             & 67.6 & 119.0 & 208.0 & --  \\
%%%% scaled virial Mass
             & 4.86 & 8.95 & 15.3 & 23.3 & 35.4 
             & 59.3 & 104.4 & 182.5 & --  \\
   $\bar{M}_{\rm h}/10^{12}~[M_\odot/h]$ 
%%%%  FOF Mass 
%             & 2.96 & 4.65 & 7.08 & 9.37 & 14.7 
%             & 21.8 & 32.1 & 46.3 & 70.3 \\
%%%% scaled virial Mass
             & 2.60 & 4.08 & 6.21 & 8.22 & 12.9 
             & 19.1 & 28.2 & 40.6 & 61.7 \\
   $\bar{n}_{\rm h}/10^{-4}~[h^3{\rm Mpc}^{-3}]$       
             & 15.7  & 12.6 & 9.46 & 6.87 & 4.87 
             & 3.47 & 2.43 & 1.64 & 1.09  \\
   $\bar{n}_h P_{\rm hh}(k=0.1~h{\rm Mpc}^{-1})$ 
             & 7.35 & 6.78 & 5.90 & 4.97 & 4.16 
             & 3.57 & 3.06 & 2.52 & 2.12 \\
   $\bar{n}_h P_{\rm hh}(k=0.2~h{\rm Mpc}^{-1})$ 
             & 2.71 & 2.54 & 2.22 & 1.88 & 1.58 
             & 1.34 & 1.14 & 0.93 & 0.76 \\
   $b_1^{\rm eff}$ (from $P_{{\rm hm}}/P_{m}$)
             & 1.06 & 1.13 & 1.22 & 1.32 & 1.44 
             & 1.60 & 1.78 & 1.99 & 2.27 \\
   ${\mathcal C}$ (from SPT)
%             & 0.67 &  0.58 & 0.41 & 0.26 & 0.12 
%             & -0.03 & -0.14 & -0.21 & -0.29 \\
             & 0.66 &  0.58 & 0.42 & 0.27 & 0.13 
             & 0.00 & -0.14 & -0.21 & -0.29 \\
   ${\mathcal C}$ (from CPT)
%             & 0.74 &  0.64 & 0.47 & 0.30 & 0.15 
%             & 0.01 & -0.10 & -0.17 & -0.24 \\
             & 0.74 &  0.65 & 0.48 & 0.32 & 0.17 
             & 0.04 & -0.10 & -0.17 & -0.24 \\
   $\bar{b}_1(M_h)$
%             & 1.18 & 1.28 & 1.38 & 1.49 & 1.63 
%             & 1.82 & 2.02 & 2.23 & 2.60 \\
             & 1.17 & 1.25 & 1.34 & 1.44 & 1.55 
             & 1.69 & 1.91 & 2.10 & 2.33 \\
   $\bar{b}_2(M_h)$
%             & -0.47 & -0.44 & -0.40 & -0.32 & -0.19 
%             & 0.078 & 0.47  & 0.96 & 2.43 \\
             & -0.47 & -0.45 & -0.42 & -0.36 & -0.27 
             & -0.12 & 0.22  & 0.60 & 1.19 \\
  \hline\hline
  \end{tabular}
  \end{center}
  \caption[simulation summary]{Summary of the catalogs of simulated halos, built
    from the $N$-body simulation outputs at $z=0.35$ for $\Lambda$CDM model
    (see text for details). We use the 9 halo catalogs, named as ``bin 1'',
    ..., ``bin 9'', which are different in their mass bins; $M_{\rm min}$
    and $M_{\rm max}$ are the minimum and maximum masses to define each mass
    bin. 
    $\bar{M}_h$ and $\bar{n}_{h}$ are the average halo virial masses
    and the mean number density of halos in each halo catalog,
    respectively. As an indicator of the shot noise contamination to the
    halo power spectrum, we give the values of $\bar{n}_{\rm h} P_{\rm hh}$
    at $k=0.1$ and $0.2~h{\rm Mpc}^{-1}$, respectively, measured from the
    simulations.  The quantity $b_1^{\rm eff}$ is the renormalized linear
    bias parameter, which is estimated by comparing the PT model prediction
    to  the simulation result for the
    halo-matter spectrum, $P_{{\rm h}m}(k)$, at
    large scales ($k<0.05 h{\rm Mpc}^{-1}$) for each halo catalog (see text for
    details). The quantity ${\mathcal C}$ is the best-fit parameter to
    characterize the residual shot noise, estimated  by fitting the PT model to
    the simulation
    result for the halo power spectrum, $P_{\rm hh}(k)$,
    in the weakly nonlinear regime (see text).  The values $\bar{b}_1(M_h)$
    and $\bar{b}_2(M_h)$ are the halo bias parameters averaged by the halo
    mass function over the halo mass range. The effective bis parameter
    $b_1^{\rm eff}$ differs from the halo bias $\bar{b}_1(M_h)$ by about
    10\% for the halo masses we consider. 
    \label{tab:sim}}
\end{table*}
%
%
%=====================================================================================
\subsection{$N$-body simulations and the halo catalogs}
\label{ssec:simulation}
%=====================================================================================

To test the accuracy of our method for modeling the halo-matter and
halo-halo power spectra in the weakly nonlinear regime, we use
$N$-body simulations for $\Lambda$CDM model done in
\citet{NishimichiTaruya:2011}. In brief, we adopted $1280^3$ $N$-body
particles and the box size of volume $1.5~({\rm Gpc} h^{-1})^3$, and used
the simulation outputs at $z=0.35$.  We defined halos using the
Friends-of-Friends (FoF) finder algorithm with linking length 0.2
times the mean particle separation. For each halo, we use the total
mass of member $N$-body particles as the halo mass, i.e. the FoF mass,
and the center-of-mass positions of the particles as the halo
position.  Then we computed the halo power spectrum from the discrete
distribution of halos in each simulation realization using the
cloud-in-cells interpolation method and Fourier transformation. To
reduce the statistical scatters, we use the mean spectra from 15
realizations.  To explore the validity of our model for different
ranges of halo masses, we divided the halos into different mass
bins. Table~\ref{tab:sim} shows the parameters of the halo catalogs;
we consider the halo catalogs divided into 9 mass bins, with mean
masses ranging from $2.96\times 10^{12}$ to $7.03\times
10^{13}~M_\odot/h$.  Note that, for the following results, we use the
shot-noise-subtracted halo spectra, where we subtracted the
theory-expectation $1/\bar{n}_{\rm h}$ from the halo spectra measured
from the simulations ($\bar{n}_h$ is the mean number density of halos
in a given simulation).

The halo bias and the halo mass function we use are given as a
function of the virial mass, rather than the FoF mass. We use the
conversion relation $M_{\rm vir}=0.88M_{\rm FoF}$ at $z=0.35$ to
estimate the average virial mass for each mass bin using the method in
\citet{HuKravtsov:03}, where we assume that the FoF halo mass is close
to the enclosed mass $M_{180b}$ \citep{White:02} inside which the mean
density is 180 times the mean mass density and also assume that each
halo follows an Navarro-Frenk-White profile \citep{NFW} with
concentration parameter $c_{\rm vir}=4$.  We should note that this
mass conversion only slightly changes the model predictions for the
halo spectra, by up to a few \% in the amplitudes.

To compute the power spectrum for halos in the finite mass range used
for the halo catalogs in Table~\ref{tab:sim}, we use the following HOD
in on our model (Eqs.~\ref{eq:pgm_renorm} and \ref{eq:pgg_renorm}):
\begin{equation}
N_{\rm HOD}(M)=
\left\{
\begin{array}{ll}
1 & \mbox{if } M_{{\rm min}, i}\le M\le M_{{\rm max},i}, \\
0 & \mbox{otherwise}. \\
\end{array}
\right.
\label{eq:HOD}
\end{equation}
%

%=====================================================================================
\subsection{Halo-matter cross-power spectrum}
\label{sssec:halo-mass-spectra}
%=====================================================================================
\begin{figure}
  \includegraphics[width=\linewidth]{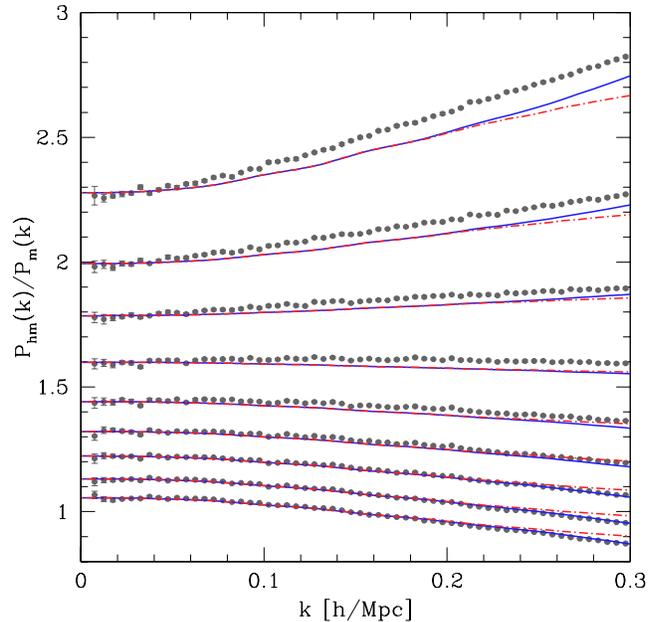} 
  \caption[cross spectrum bias]{ The bias function $b^{\rm cross}_{\rm
      h}(k)$, defined as $b^{\rm cross}_{\rm h}(k)\equiv P_{{\rm
        h}m}(k)/P_m(k)$ for halos of different mass ranges given in
    Table~\ref{tab:sim}.  The symbols are the simulation results,
    which are computed from 15 simulation realizations (see text for
    details), and clearly show a scale-dependent bias at $k\simgt
    0.1~h{\rm Mpc}^{-1}$.  The red dot-dashed curves show the PT predictions
    including up to the one-loop correction (Eqs.~\ref{eq:phm_renorm}
    and \ref{eq:pgm_renorm}). We determined the free parameter of the
    model prediction, the effective linear bias parameter $b^{\rm
      eff}_1$, by fitting the model prediction to the simulation
    result up to $k\le 0.05~h{\rm Mpc}^{-1}$ for each halo mass bin.  The
    scale-dependent bias is from a combination of the halo bias
    ($b_2$) and the nonlinear matter power spectrum.  The solid curves
    show the model predictions when using the improved PT model
    prediction given in \citet{NishimichiTaruya:2011} for the
    nonlinear matter power spectrum $P^{\rm NL}_m$ instead of the
    standard PT. \label{fig:p_hm} }
\end{figure}

\begin{figure}
  \includegraphics[width=\linewidth]{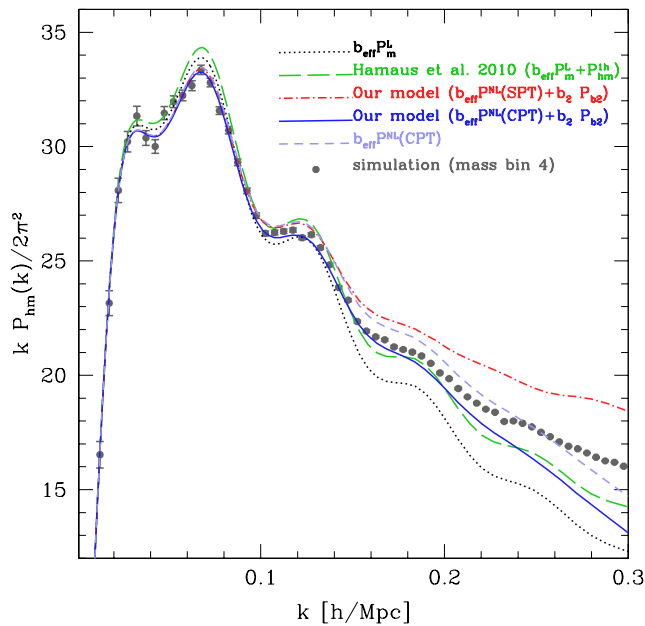} 
  \caption[halo matter cross spectrum]{ 
    The halo-matter cross-power spectrum $P_{{\rm h}m}(k)$ for
    halos of mass bin ``4'' (Table~\ref{tab:sim}), plotted in the unit of
    $kP_{{\rm h}m}(k)/2\pi^2$ so that the features in the weakly nonlinear
    regime including the BAO features become prominent. The symbols are
    the simulation result as in Fig.~\ref{fig:p_hm}, while the error bars
    are the statistical uncertainties at each $k$ bins estimated from the
    15 realizations.  The dotted-dashed curve denotes the model prediction
    (Eq.~\ref{eq:phm_renorm}) obtained by using the standard PT to compute 
    the nonlinear matter power spectrum $P_m(k)$ including up to  one-loop 
    corrections. The solid curve is the result if using the improved PT 
    prediction (CPT) for $P_m(k)$, which shows an improved agreement with 
    the simulation result up to the higher $k$. 
    To see the effect of scale-dependence bias arising from $b_2$, the short 
    dashed curve shows $b_1^{\rm eff}P_m^{\rm CPT}(k)$, i.e the nonlinear matter 
    power spectrum (CPT) multiplied by the effective linear bias parameter. 
    The dotted curve is the linear theory prediction, $b_1^{\rm eff}P^{\rm L}_m(k)$.  
    For comparison, we also show the model prediction recently
    proposed in \citet{Hamausetal:11}, which models the halo-matter power
    spectrum fully based on the halo model: the linear theory plus the
    1-halo term given as 
    $b_1^{\rm eff}P^{\rm L}_m(k)+P^{1h}(k)$.
    \label{fig:kp_hm}}
\end{figure}

First, we study the halo-matter cross-power spectrum and define the bias
function, for convenience of the following discussion, as
\begin{equation}
b_{{\rm h}_i}^{\rm cross}(k)\equiv \frac{P_{{\rm h}_i m}(k)}
{P_{m}(k)}, 
\label{eq:bk}
\end{equation}
where $P_{{\rm h}_im}$ is the cross-spectra between dark matter
($N$-body) particles and halos of the $i$-th mass bin, and $P_m$ is
the matter power spectrum computed from the original $N$-body
simulations.  The shot noise is negligible for the cross-spectrum
$P_{{\rm h}_im}$.

The data points in Fig.~\ref{fig:p_hm} show the bias function $b^{\rm
  cross}_{{\rm h}_i}(k)$ measured from the simulations. The spectra
$P_{{\rm h}_im}$ and $P_m$ in each simulation share the same
large-scale structure, and therefore the scatters due to the sampling
variance mostly cancel in the ratio, yielding relatively
smoothly-varying data points over $k$. The figure clearly shows that
the halo bias has greater amplitudes for more massive halos. At
sufficiently large scales or small $k$, $k\simlt 0.08~h{\rm Mpc}^{-1}$,
the halo bias appears to be constant, implying that the linear bias
model is valid at the large scales. On the other hand, at the larger
$k$, the simulation results manifest a scale-dependent halo bias for
all the halo mass bins.

The dot-dashed curves show our model predictions computed using
Eq.~(\ref{eq:phm_renorm}). To compute the model predictions, we need
to fix one free parameter, the renormalized linear bias $b_1^{\rm
  eff}$, for which we determined $b_1^{\rm eff}$ by fitting the
model-predicted $b^{\rm cross}(k)$ to the simulation $b^{\rm
  cross}(k)$ in the linear regime, at $k\le 0.05~h{\rm Mpc}^{-1}$.
Table~\ref{tab:sim} shows that the best-fit $b^{\rm eff}_1$ differs
from the linear halo bias by about 10\%, which is also found by
\citet{Maneraetal:10}.  In our method, we interpret that the
discrepancy between $b_1$ and $b_1^{\rm eff}$ arises due to the
renormalization; $b_1^{\rm eff}$ has a contribution of the
higher-order moments (see around Eq.~\ref{eq:phm_renorm_step}) in
addition to $b_1$.  However, the discrepancy might also be ascribed
partly to the inaccuracy of the analytical halo mass function
(Eq.~\ref{eq:dndm}), compared with our $N$-body simulations, as well
as to violation of the universality of the mass function.
\citep[e.g.][]{Tinkeretal:08}.  However, exploring these issues is
beyond the scope of this paper, so we leave these for future work.
Besides this free parameter, we used the input $\Lambda$CDM model
parameters and halo mass range to compute the model prediction.  Once
the parameter $b_1^{\rm eff}$ is determined, our model can be in
remarkably good agreement with the simulation results, including the
$k$-dependence and the halo mass-dependence.  For the largest mass
bin, our model shows a sizable disagreement, possibly due to an
inaccuracy of the halo mass function used for the model calculation or
the break-down of perturbation theory to describe too strong nonlinear
bias.

At large $k$, the perturbation theory ceases to be accurate, and is
indeed not accurate enough up to $k\sim 0.2~h{\rm Mpc}^{-1}$. There have been
many efforts to improve the PT prediction of nonlinear matter power
spectrum by including the higher-order loop corrections, e.g. the
renormalized PT \citep[][also see references
therein]{CrocceScoccimarro:06,Taruyaetal:09}.  The solid curve shows
the results if we use the improved PT prediction for the nonlinear
matter power spectrum $P^{\rm NL}_m$ that is taken from
\citet{NishimichiTaruya:2011} \citep[also see][]{Taruyaetal:09} (the
closure theory; hereafter CPT), instead of the standard PT including
up to the one-loop correction.  The improved $P_{m}^{\rm NL}$ has
smaller amplitudes in the weakly nonlinear regime $k\simgt 0.1~h{\rm Mpc}^{-1}$ at the redshift $z=0.35$ than the standard PT predicts,
yielding a slightly stronger scale-dependence of $b(k)(=P_{{\rm
    h}m}/P_m^{\rm NL})$ as evident from Eq.~(\ref{eq:phm_renorm}). The
improved $P_{m}^{\rm NL}$ shows a similar-level agreement with the
simulation results.

In order to see the accuracy of our model for $P_{{\rm h}m}(k)$, we
compare the simulation result for $P_{{\rm h}m}$ with the different
model predictions in Fig.~\ref{fig:kp_hm}.  Here we considered the
intermediate halo mass bin (bin $4$), and we use the parameter
$b_1^{\rm eff}$ determined in Fig.~\ref{fig:bias}. Encouragingly, the
figure clearly shows that our model prediction well agrees with the
simulation $P_{{\rm h}m}(k)$ up to $k\simeq 0.2~h{\rm Mpc}^{-1}$, if we
use the improved PT model (CPT) for the nonlinear matter power
spectrum. If we use the standard PT theory instead, the agreement is
only up to $k\simeq 0.12~h{\rm Mpc}^{-1}$, and the standard PT breaks down
at the larger $k$, in the weakly nonlinear regime. Hence we conclude
that the apparent agreement of the standard PT up to the higher $k$
bins in Fig.~\ref{fig:bias} is due to a cancellation of inaccuracies
in the two spectra $P_{{\rm h}m}$ and $P_{m}$, in the numerator and
denominator of the bias function $b^{\rm cross}(k)$. Comparing the
solid and short-dashed curves shows the effect of nonlinear
scale-dependent bias that arises from the term proportional to
$b_2(M)$ in Eq.~(\ref{eq:phm_renorm}). The scale-dependent bias
becomes important at $k\simgt 0.1~h{\rm Mpc}^{-1}$ for this redshift
$z=0.35$, and our model can reproduce the simulation result.

The dashed curve shows the model prediction recently proposed in
\citet{Hamausetal:11}, where the nonlinear power spectrum is computed
based on the halo model in combination with the halo bias parameters.
The figure shows that the agreement is not as good as our model
prediction.

%=====================================================================================
\subsection{Halo auto-power spectrum}
%=====================================================================================
%
\begin{figure}
  \includegraphics[width=\linewidth]{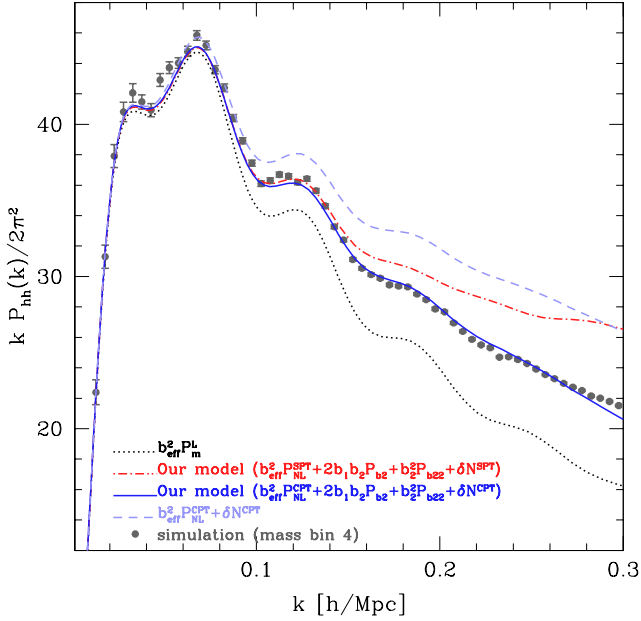} 
  \caption{
    As in Fig.~\ref{fig:kp_hm}, but for 
    the halo auto-power spectrum $P_{\rm hh}(k)$.
    The symbols show the simulation result for the halo catalog of
    mass bin $4$, where the standard shot noise term $1/\bar{n}_{\rm h}$ is subtracted
    from the measured power spectrum.  
    The dot-dashed and solid curves are the model predictions
    (Eq.~\ref{eq:phh_renorm}), where we used the standard PT and the
    improved PT (CPT) to compute the nonlinear matter power spectrum $P^{\rm NL}_m(k)$, 
    respectively. To compute the model predictions, we used the
    same linear bias parameter $b_1^{\rm eff}$ in Fig.~\ref{fig:p_hm}, and
    determined the residual shot noise parameter $\delta N$ from the
    fitting of each model prediction to the simulation result up to
    $k=0.2~h{\rm Mpc}^{-1}$.
    To illuminate the scale-dependent bias, the
    dashed curve shows the model prediction ignoring the terms of halo
    bias parameters in Eq.~(\ref{eq:phh_renorm});
    i.e.  $(b_1^{\rm eff})^2 P_m^{\rm CPT}(k)+\delta N$. 
    The dotted curve is the linear theory prediction, 
    $(b_1^{\rm eff})^2P^{\rm L}_m$. 
    \label{fig:kp_hh}}
\end{figure}
\begin{figure}
  \includegraphics[width=\linewidth]{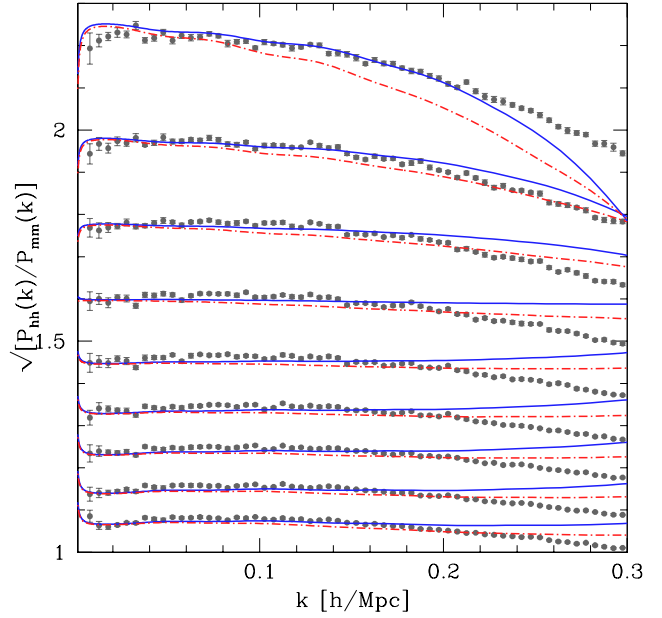} 
  \caption{
    The bias function defined in terms of the halo auto-spectrum as 
    $b^{\rm auto}(k)=\sqrt{P_{\rm hh}(k)/P_{m}(k)}$ for each halo 
    mass as in Fig.~\ref{fig:p_hm}.
    The linear theory predicts no scale dependence and that the bias 
    amplitude is the same to that of large scale limit for 
    $P_{{\rm h}m}(k)/P_m(k)$ in Fig.~\ref{fig:p_hm}. The symbols are the 
    simulation results for different halo masses, where the standard 
    shot noise $1/\bar{n}_{\rm h}$ is subtracted.
    To compute the model predictions, we need to fix the free parameters: 
    we used the same  $b_1^{\rm eff}$ to that in Fig.~\ref{fig:p_hm} and  
    determined the residual shot noise parameter ${\mathcal C}$ by fitting 
    the model prediction to the simulation result up to 
    $k\le 0.2~h{\rm Mpc}^{-1}$.
    \label{fig:p_hh}}
\end{figure}

The model parameters for the halo power spectrum $P_{\rm hh}(k)$ are
$b_1^{\rm eff}$ and the residual shot noise parameter $\delta N$.  For
convenience of our discussion, we use the following parametrization of
the residual shot noise term relative to the standard shot noise term:
\begin{equation}
  \delta N_{h_i}={\mathcal C}_i\frac{1}{\bar{n}_{h_i}}. 
\end{equation}

Fig.~\ref{fig:kp_hh} compares the PT predictions and the simulation
results for $P_{\rm hh}(k)$, as in Fig.~\ref{fig:kp_hm}. To compute
the PT predictions, we determined the free parameters $b_1^{\rm eff}$
and ${\mathcal C}$ for each halo catalog as follows. For $b_1^{\rm
  eff}$, we used the same values as those used for the halo-matter
cross-power spectra in Fig.~\ref{fig:p_hm}. For ${\mathcal C}$, we
determined the value by fitting the model prediction to the simulation
result up to $k_{\rm max}=0.15$ or $0.20~h{\rm Mpc}^{-1}$ for the SPT or
the improved PT (CPT), respectively. The maximum wavenumber $k_{\rm
  max}$ is chosen because each of the PT models for nonlinear {\em
  matter} power spectrum is accurate enough up to the $k_{\rm
  max}$-wavenumber to within about 3\% accuracy
\citep{NishimichiTaruya:2011}. In doing this fitting, we accounted for
the statistical uncertainties in estimating the power spectrum from
the 15 simulation realizations; i.e., we used the weighting in each
$k$-bin, given as $(\Delta P_{\rm hh})^2\propto 1/(2\pi k_i^2\Delta
k)[P_{\rm hh}(k_i)+1/\bar{n}_{\rm h}]^2$ ($k_i$ is the central value
of the $i$-th $k$-bin and $\Delta k$ is the width).  The best-fit
residual shot noise parameter ${\mathcal C}$ is about 30\% compared to
the standard shot noise for this halo mass bin (bin 4).
Table~\ref{tab:sim} shows a strong anti-correlation between ${\mathcal
  C}$ and halo mass. The anti-correlation might be ascribed to a mass
dependence of the stochastic halo bias
\citep{Matsubara:99,TaruyaSuto:00}.

As can be found from Fig.~\ref{fig:kp_hh}, our model prediction is in
{\em good} agreement with the simulation result, apparently up to
$k\simeq 0.25~h{\rm Mpc}^{-1}$, {\em if} we use the improved PT prediction
(solid curve) \footnote{Note that, exactly speaking, the improve PT
  (CPT) ceases to be accurate at $k>0.2~h{\rm Mpc}^{-1}$, so the apparent
  agreement at the scales is as a result of the residual shot noise
  contribution, which happens to match the simulation result.}.  The
nice agreement is found by including the residual shot noise
contribution, which can account for a part of the nonlinear bias
effect.  The standard perturbation theory cannot achieve the similar
level agreement, even if varying the residual shot noise parameter.
The figure also shows other model predictions, and the comparison of
different model predictions manifests the importance of nonlinear
clustering effect and scale-dependent bias in the weakly nonlinear
regime.  Combining the results in Figs.~\ref{fig:kp_hm} and
\ref{fig:kp_hh} implies that the halo bias parameters $b_1(M)$ and
$b_2(M)$, in combination with the nonlinear matter power spectrum and
the residual shot noise, can well reproduce the halo spectra $P_{{\rm
    h}m}(k)$ and $P_{\rm hh}(k)$.

Now we study another bias function defined in terms of the halo power
spectrum as
\begin{equation}
  \label{eq:bias_est_HM}
  b^{\rm auto}_{h_i}(k)
  \equiv 
\sqrt{\frac{P_{h_i h_i}(k)}{P_{m}(k)}}.
\end{equation}
This bias function is different from the bias function $b^{\rm
  cross}(k)$ we studied in Fig.~\ref{fig:bias}, as can be explicitly
found from Eqs.~(\ref{eq:phm_renorm}) and (\ref{eq:phh_renorm}) (also
Eqs.~\ref{eq:pgm_renorm} and \ref{eq:pgg_renorm}).
Fig.~\ref{fig:p_hh} compares the PT predictions and the simulations
results for the bias function of each halo catalog $b^{\rm
  auto}_{h_i}(k)$, where the PT predictions are computed by using the
best-fit power spectra $P_{{\rm h}m}$ and $P_{\rm hh}$ as estimated in
Figs.~\ref{fig:p_hm} and \ref{fig:kp_hh}.  Note that the model in
Fig.~\ref{fig:p_hh} appears to show a larger disagreement with the
simulation result at $k>0.2 h{\rm Mpc}^{-1}$, compared to Fig.~\ref{fig:kp_hh},
especially when using the CPT for $P_m^{\rm NL}$.  This is mainly
because of the unphysical damping of the CPT model for the $P^{\rm
  NL}_m$ at that scales \citep{Taruyaetal:09}, and partly because the
plotting range of $y$-axis in Fig.~\ref{fig:p_hh} is narrower than in
Fig.~\ref{fig:kp_hh}. We should notice that the discrepancy of the
model from the simulation is less than $1\%$ at $k<0.2~h{\rm Mpc}^{-1}$ for 
both the plots.  The best-fit value of
the residual shot noise parameter, ${\mathcal C}$, for each halo mass
bin is given in Table~\ref{tab:sim}. The amount of the residual shot
noise varies with halo masses, ranging from a few \% to 70\% compared
to the standard shot noise term, and changes from negative to positive
values from less to more massive halos.  The bias function $b^{\rm
  auto}(k)$ shows a scale-dependence over the range of $k$ we
consider. The scale-dependence of $b^{\rm auto}$ indeed differs from
that of $b^{\rm cross}(k)$ in Fig.~\ref{fig:p_hm} as the PT model
predicts.  The model predictions can fairly well reproduce the
simulation results up to $k\simeq 0.2h{\rm Mpc}^{-1}$, but then show a
larger disagreement at the larger $k$ than in Fig.~\ref{fig:p_hm} due
to inaccuracies in both the model predictions for $P_{{\rm h}m}$ and
$P_{\rm hh}$.

%=====================================================================================
\subsection{Comparison of halo bias model with other models}
\label{ssec:halo_model_comparison}
%=====================================================================================
%
\begin{figure*}
  \includegraphics[width=\linewidth]{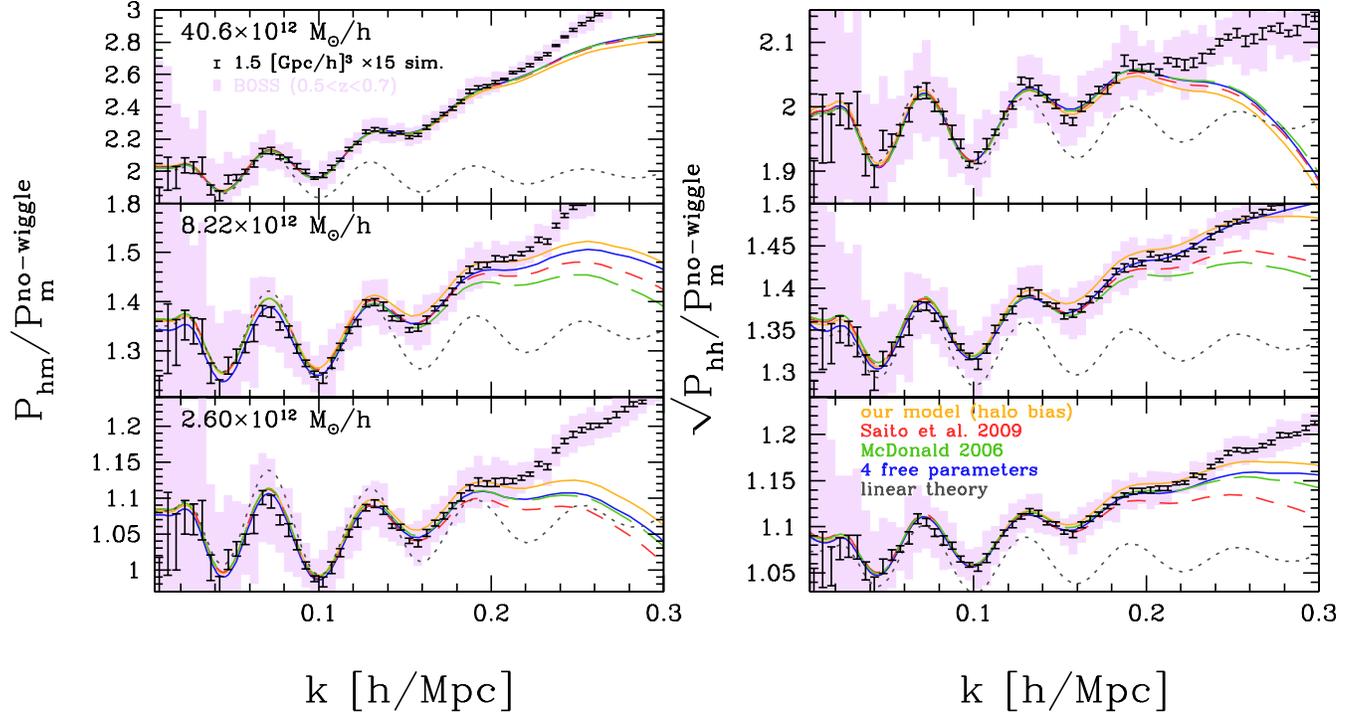}
  \caption{ Comparison of the different PT-based model predictions for
    the halo-matter cross-power spectrum ({\em left panel}) and the
    halo auto-power spectrum ({\em right panel}), for the three halo
    catalogs of different mass bins (the 1, 4 and 8 mass bins in
    Table~\ref{tab:sim}). For illustrative purpose, the power spectra
    are normalized by the nonlinear matter power spectrum without BAO
    wiggles. For comparison, the dotted curves are the linear theory
    prediction (the power spectra in the denominator and numerator are
    both the linear power spectra).  The different models are
    expressed by the similar forms (Eq.~\ref{eq:nlps_app}), and have a
    different number of free parameters (2, 3 and 4 parameters), as
    indicated by legends and explained below
    Eq.~(\ref{eq:nlps_app}). The data with error bars at each $k$-bin
    are the spectra estimated from 15 realizations each of which has a
    1.5~(Gpc$ h^{-1})^3$ volume, and the error bar is the $1\sigma$ scatter
    of the central value at each $k$-bin, which is estimated by
    dividing the scatters of 15 realizations by $\sqrt{15}$;
    i.e. $1\sigma$ statistical scatter for a volume of $15\times
    1.5~({\rm Gpc} h^{-1})^3$. The best-fit parameters for each model are
    obtained by comparing the model predictions to the simulation
    taking into account the statistical errors
    (Eq.~\ref{eq:chi2}). The shaded region at each $k$ bin is the
    errors expected for a BOSS-like survey of 3.4 (Gpc$ h^{-1})^3$ volume,
    obtained by assuming that the error bars scales with survey volume
    ($V_s$) as $\sigma(P)\propto 1/\sqrt{V_s}$. Our model and the
    other models show a similar-level agreement with the simulations,
    to within the $1\sigma$ error bars of BOSS-like survey up to
    $k\simeq 0.2~h{\rm Mpc}^{-1}$, for different halo mass bins.
    \label{fig:ps_bestfit}}
\end{figure*}

The main difference between our method and the previous works
\citep{Mcdonald:2006,JeongKomatsu:09,Saitoetal:09,baldaufetal:2010,Saitoetal:11},
is whether or not to incorporate the renormalized bias approach and
the halo bias into the PT approach for computing the nonlinear power
spectra. In this section, we compare our model predictions with other
models that have similar forms.

The nonlinear power spectra in these models are expressed by the
following general forms (see Eqs.~\ref{eq:phm_renorm} and
\ref{eq:phh_renorm} for our model):
\begin{align}
  \label{eq:nlps_app}
  P_{{\rm h}m}(k)
  &=
  \alpha_1 
  P^{\rm NL}_m(k) + \alpha_2 P_{b2}(k), \nonumber\\
  P_{\rm hh}(k)
  &=
  \alpha_1^2  P^{\rm NL}_m(k)
  +2 \alpha_3 P_{b2}(k) + \alpha_2^2 P_{b,22}
  + \alpha_4,
\end{align}
where 
\begin{align}
  P_{b2}(k)
  &\equiv
  \int
  \frac{\id^3\bs{q}}{(2\pi)^3}\! P_m(q)P_m(|\bs{k}-\bs{q}|)
  F_{2}(\bs{q},\bs{k}-\bs{q}),\nonumber\\
  P_{b,22}(k)
  &\equiv
  \frac{1}{2}\int\! \frac{\id^3\bs{q}}
  {(2\pi)^3} 
  \left[
    P_m(q)P_m(|\bs{k}-\bs{q}|) 
    -P_m(q)^2
  \right].
\end{align}
In terms of these forms, we can categorize the different models as
\begin{itemize}
 \item {\em Our model}: $\alpha_1 = b_1^{\rm eff}$,
$\alpha_2 = b_2(M)$, $\alpha_3 = b_1(M) b_2(M)$, $\alpha_4 = \delta N$,
       where $b_1(M)$ and $b_2(M)$ are the halo bias parameters and 
$b_1^{\rm eff}$ and $\delta N$ are treated as free
       parameters. 
\item {\em Saito et al. 2009}: This is based on a renormalized
      perturbation theory originally proposed in
      \citet{Mcdonald:2006}. Here the coefficients are set to 
$\alpha_1=b_1$, $\alpha_2=b_2$,
      $\alpha_3=b_1 b_2$, and $\alpha_4=\delta N$, and the three 
($b_1,b_2,\delta N$) are treated
      as free parameters to be determined by the fitting. 
This method is also studied in
      \citet{Baldaufetal:10}. 
\item {\em McDonald 2006}: This is similar to the method ``Saito et
      al. 2009'', but uses the nonlinear matter power spectra $P^{\rm
      NL}_m$ for $P_m$'s in $P_{b2}$ and $P_{b,22}$, instead of the
      linear spectrum. This method is
       intended
      to include renormalization for bias parameters as well as for the
      nonlinear matter power spectrum. 
\item {\em 4 free parameters}: This is a variant of our model. The
      coefficients  are set to $\alpha_1 = b_1^{\rm eff}$,
$\alpha_2 = b_2$, $\alpha_3 = b_1 b_2$, $\alpha_4 = \delta N$, 
and all the 4 parameters ($b_1^{\rm eff}$, $b_1$, $b_2$, $\delta N$)
      are treated as free parameters. 
\end{itemize}
Note that, to have a fair comparison, we use CPT to compute $P^{\rm
  NL}_m(k)$ in the first term of Eq.~(\ref{eq:nlps_app}) for all the
above models.  Thus the different models have different ranges of
their variations in the power spectra as a function of $k$ with
varying free parameters for a given cosmological model.

Fig.~\ref{fig:ps_bestfit} shows the different model predictions,
described above, for $P_{{\rm h}m}$ and $P_{\rm hh}$ for different
halo mass bins, compared to the simulation results. To determine the
free parameters in each model, we minimize the following $\chi^2$ by
comparing the model prediction to the simulation result:
\begin{equation}
\chi^2
  =
  \sum
  \left( \bs{P}^{\rm sim} - \bs{P}^{\rm model} \right)^T
  {\bs{\rm C}}^{-1} 
  \left( \bs{P}^{\rm sim} - \bs{P}^{\rm model} \right),
\label{eq:chi2}
\end{equation}
where $\bs{P}(k_i)=[P_{{\rm h}m}(k_i), P_{\rm hh}(k_i)]$, the power
spectra with superscripts ``sim'' or ``model'' are the simulated or
model power spectra, respectively, {\bf C} is the covariance matrix of
the power spectrum computed from 15 realizations of the simulated
power spectra, and {\bf C}$^{-1}$ is the inverse matrix. We use the
power spectrum information up to $k_{\rm max}=0.2~h{\rm Mpc}^{-1}$. We
included correlation between $P_{{\rm h}m}$ and $P_{\rm hh}$ at the
same $k$-bin in the covariance matrix, but ignored correlations
between different $k$ bins for simplicity \footnote{This assumption
  would not cause any serious systematic errors as the scales we
  consider are in the weakly nonlinear regime and the non-Gaussian
  errors, which cause correlations between different $k$ bins, are not
  significant as studied in \citet{Takahashietal:09}.}. Our choice of
$k_{\rm max}=0.2~h{\rm Mpc}^{-1}$ is based on the fact that the CPT
prediction for $P^{\rm NL}_m(k)$ is accurate to within a 3\% level up
to $k=0.2~h{\rm Mpc}^{-1}$ compared to the simulated spectrum at $z=0.35$
as shown in \citet{NishimichiTaruya:2011}.

Fig.~\ref{fig:ps_bestfit} shows that the different models well
reproduce the simulated $P_{{\rm h}m}$ or $P_{\rm hh}$ to an
equal-level accuracy up to $k\simeq 0.15$ or $0.2~h{\rm Mpc}^{-1}$ in some
cases. Again note that our model has least free parameters among these
models, because our model uses the halo bias parameters $b_1(M)$ and
$b_2(M)$ to model the scale-dependent bias of the nonlinear halo power
spectra (Eqs.~\ref{eq:nlps_app}), and therefore restricts a range of
the model variations compared to other models. Nevertheless, our model
appears to be reasonably accurate compared to other models.

For comparison, the shaded region around the curves in
Fig.~\ref{fig:ps_bestfit} shows $1\sigma$ statistical errors of the
power spectrum measurements expected for a survey with volume coverage
of about $3.4~({\rm Gpc} h^{-1})^3$, which roughly corresponds to the
volume of a BOSS-like survey with redshift range $0.5\le z\le 0.7$ and
area coverage $10,000$ square degrees. We estimated the error bars by
scaling the scatters at each $k$ bin from the 15 simulation
realizations assuming that the scatters scale with a survey volume as
$\sigma(P)\propto 1/\sqrt{V_s}$.  It can be found that our model is
accurate at least enough up to $k\simeq 0.20~h{\rm Mpc}^{-1}$ within the
$1\sigma$ errors for a BOSS-like survey, for a different range of
halos masses. The linear theory is not accurate at BAO scales, at
$k\simgt 0.1~h{\rm Mpc}^{-1}$, although the PT based model also ceases to
be accurate at $k\simgt 0.2~h{\rm Mpc}^{-1}$.

\begin{table*}
    \begin{tabular}{lccccccccc}\hline\hline
      Model            & bin 1 & bin 2 & bin 3 & bin 4 & bin 5 & bin 6 & bin 7 & bin 8 & bin 9 \\ \hline
      This work        & 0.38(0.54) &  0.52(0.75) &  0.51(0.75) &  0.62(0.86) &  0.53(0.70) &  0.60(0.76) &  0.32(0.34) &  0.23(0.33) &  0.46(1.3) \\
      Saito et al.2009 & 0.34(0.43) &  0.49(0.62) &  0.45(0.78) &  0.68(0.80) &  0.67(0.61) &  0.30(0.31) &  0.32(0.35) &  0.23(0.26) &  0.50(0.95) \\
      McDonald 2006    & 0.30(0.32) &  0.51(0.51) &  0.50(0.49) &  0.71(0.74) &  0.63(0.85) &  0.50(0.77) &  0.39(0.43) &  0.22(0.26) &  0.50(0.82) \\
      4 free parameters& 0.19(0.21) &  0.27(0.28) &  0.19(0.21) &  0.20(0.20) &  0.23(0.21) &  0.30(0.31) &  0.25(0.26) &  0.22(0.26) &  0.24(0.36) \\
      linear theory    & 5.1 (13.4) &  5.4 (15.7) &  4.3 (15.4) &  4.4 (15.6) &  3.6 (18.9) &  4.3 (23.3) &  8.4 (35.8) &  12.6(49.0) &  19.8(71.0) \\ \hline\hline
    \end{tabular}
\caption{Comparison of the different models as in
  Fig.~\ref{fig:ps_bestfit}, but the numbers in each row- and column are
  the reduced $\chi^2$-values ($\chi^2_\nu$) for the best-fit power
  spectra of each model.  The different columns are for the halo
  catalogs of different mass bins (see Table~\ref{tab:sim}).  To compute
  the reduced $\chi^2$ values, we obtained the best-fit model up to
  $k_{\rm max}=0.15~h{\rm Mpc}^{-1}$ by fitting the model prediction to the
  simulation spectra assuming the errors for a BOSS-like survey, while
  the value in parenthesis is the value for $k_{\rm max}=0.20~$. The
  degrees of freedom for the $\chi^2$ fitting is: 58 or 78 in total for
  $P_{{\rm h}m}$ and $P_{\rm hh}$ for $k_{\rm max}=0.15$ or 0.2~$h{\rm Mpc}^{-1}$, 
  respectively, minus the number of model parameters (either 2,
  3 or 4). Our model has 2 free parameters, the models ``Saito et
  al. (2009)'' and ``McDonald (2006)'' have 3 parameters, and the model
  ``4 free parameters'' has 4 parameters. The linear theory breaks down,
  but the different models of the nonlinear power spectra are equally
  acceptable for a BOSS-like survey.
  \label{tab:comparison}}
\end{table*}

\begin{figure}
  \includegraphics[width=\linewidth]{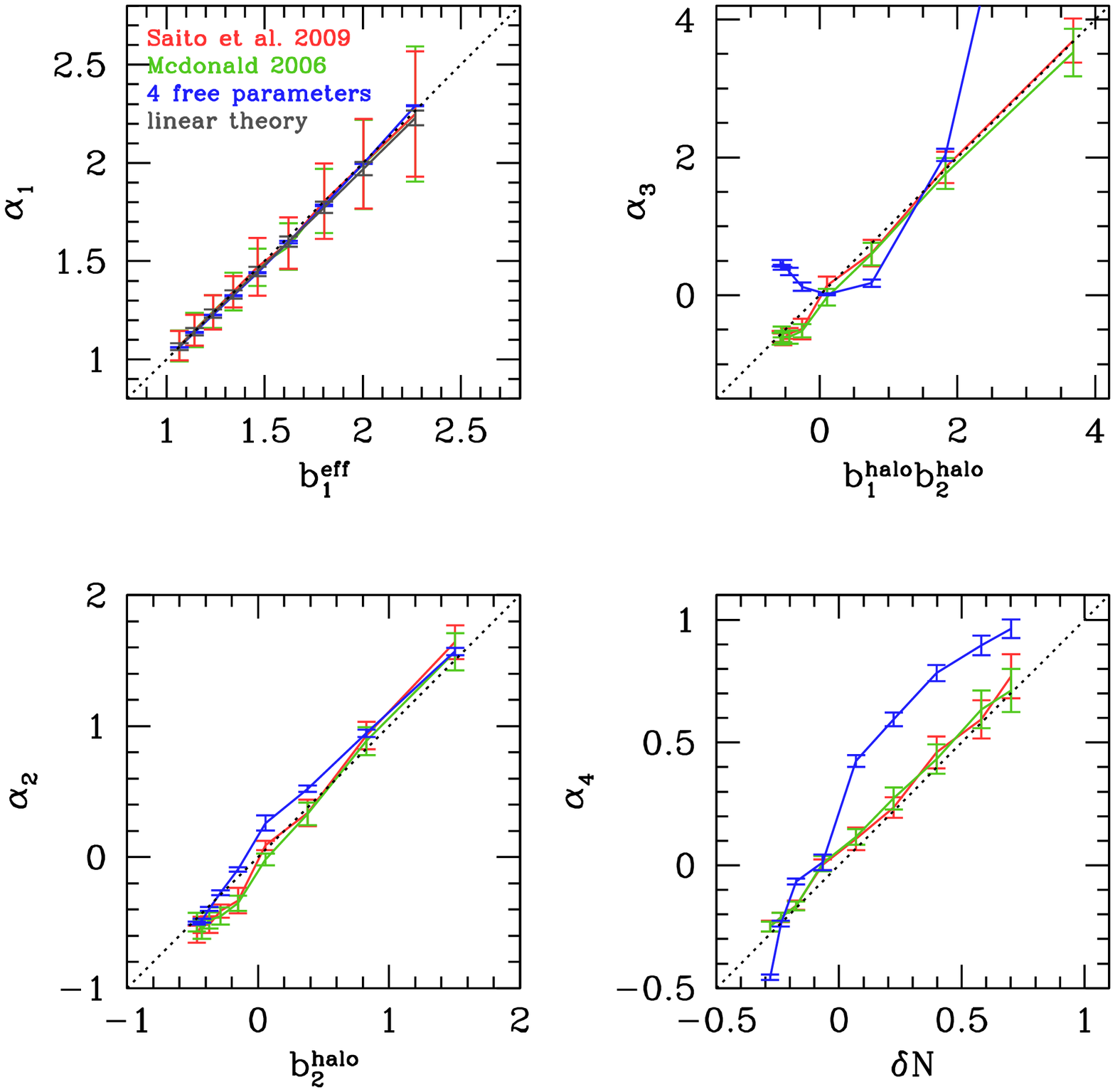}
  \caption{Comparison of the best-fit coefficients ($\alpha_1$,
 $\alpha_2$, $\alpha_3$, $\alpha_4$) in the model
 nonlinear power spectra (Eq.~\ref{eq:nlps_app}) with the best-fit
 parameters of our model, obtained from the 9 halo catalogs. 
Here we consider the different models as in
 Fig.~\ref{fig:ps_bestfit} (also see below Eq.~\ref{eq:nlps_app}).
Note that $b_1(M)$ and $b_2(M)$ in the $x$-axis of the 
 upper-right or lower-left panels are the halo bias parameters for
 different halo mass bins. The best-fit parameters are obtained by using
 the power spectra $P_{{\rm h}m}$ and $P_{\rm hh}$ up to $k=0.2~h{\rm Mpc}^{-1}$ 
 for a BOSS-like survey, as in Fig.~\ref{fig:ps_bestfit} or
 Table~\ref{tab:comparison}. The panels show that our model is
 almost equivalent to the models ``Saito et al. (2009)'' and ``McDonald
 (2006)'', implying that the halo bias parameters are a good
 approximation for the forms of the nonlinear power spectra given by
 Eq.~(\ref{eq:nlps_app}). 
 \label{fig:b2compare}}
\end{figure}

Fig.~\ref{fig:b2compare} and Table~\ref{tab:comparison} give a more
quantitative comparison of our model with other models, where we
consider the BOSS-like survey to compute the $\chi^2$ given the
expected measurement accuracies as in Fig.~\ref{fig:ps_bestfit}.
Fig.~\ref{fig:b2compare} compares the best-fit coefficients
($\alpha_1,\dots, \alpha_4$), which can be read as effective bias
parameters (see below Eq.~\ref{eq:nlps_app}), with the results of our
model; the halo bias parameters ($b_1(M), b_2(M)$) and the best-fit
parameters $b_1^{\rm eff}$ and $\delta N$. The figure shows that the
results for ``Saito et al. (2009)'' or ``McDonald 2006'' reproduce the
similar results to our model. The model ``4 free parameters'' shows a
sizable difference from our result, especially for $\alpha_3$ and
$\delta N$, implying that even small changes in the coefficients give
the similar nonlinear power spectra for $P^{\rm NL}_{{\rm h}m}$ and
$P^{\rm hh}$ at the scales. This also means a strong degeneracy
between $\alpha_3$ and $\delta N$ parameters.
Table~\ref{tab:comparison} gives the reduced $\chi^2$ of the best-fit
model power spectra for each model and for each halo mass bin.  Here
we consider $k_{\rm max}=0.15$ or $0.2~h{\rm Mpc}^{-1}$ for the maximum
wavenumber to use for the model fitting. The degrees of freedom are
defined by the number data point of the simulated spectra (58 or 78 in
total for $P_{{\rm h}m}$ and $P_{\rm hh}$ when employing $k_{\rm
  max}=0.15$ or $0.2~h{\rm Mpc}^{-1}$, respectively) minus the number of
free parameters (either 2, 3 or 4). Again our model and the models
``Saito et al. (2009)'' or ``McDonald (2006)'' are in a similar-level
accuracy, and the model ``4 free parameters'' is slightly better due
to more free parameters. Note that the reduced $\chi^2$ value is
smaller than unity, partly because the central values of the simulated
spectra are taken from the simulations of 22.5~$({\rm Gpc} h^{-1})^3$
volume and therefore the central value have smaller scatters than
expected from a BOSS-like survey.

%=====================================================================================
%=====================================================================================
\section{Summary and Discussion}
%=====================================================================================
%=====================================================================================

In this paper, we have studied a method of modeling the nonlinear halo
power spectra, $P_{{\rm hm}}(k)$ and $P_{\rm hh}(k)$, by combining the
PT approach of structure formation, the local bias ansatz and the halo
bias.  The nonlinearities of halo power spectra, which are deviations
form the linear theory prediction, arise from the two effects: the
nonlinear matter clustering and the nonlinear relation between the
matter and halo density fields. In deriving the nonlinear halo power
spectra, we employed the renormalization approach
\citep{Mcdonald:2006} to re-sum the higher-order terms so that the
terms are replaced with the nonlinear matter power spectrum,
multiplied by the ``renormalized'' linear bias parameter. The
remaining terms in the nonlinear halo spectra are given as a function
of the linear matter power spectrum and the halo bias, where the terms
at the one-loop correction order are included.  As a result, we
expressed the halo-matter cross-power spectrum
(Eq.~\ref{eq:phm_renorm}) in terms of the nonlinear and linear matter
power spectra, the halo bias ($b_2(M)$), and one free parameter, the
renormalized linear bias parameter ($b_{1}^{\rm eff}$), which needs to
be determined in the linear regime.  Similarly, the halo auto-power
spectrum (Eq.~\ref{eq:phh_renorm}) is given as a function of the
nonlinear and linear matter power spectra, the halo bias functions
($b_1(M), b_2(M)$), and the two free parameters, $b_1^{\rm eff}$ and
the residual shot noise parameter $\delta N$.  Thus our method
utilizes the recent development in an accurate model of the nonlinear
matter power spectrum based on the refined perturbation theory and/or
$N$-body simulations. In our model, the halo power spectra are
specified by cosmological parameters, halo mass, redshift, and a less
number of free parameters.

We showed that our model predictions for $P_{{\rm h}m}(k)$ and $P_{\rm
  hh}(k)$ are in nice agreement with the simulation results, up to
$k\simeq 0.2~h{\rm Mpc}^{-1}$, at simulation output $z=0.35$ (see
Figs.~\ref{fig:kp_hm} and \ref{fig:kp_hh}), if using the improved PT
theory prediction for the nonlinear matter power spectrum in the model
calculation.  The linear power spectrum breaks down at $k\simeq
0.1~h{\rm Mpc}^{-1}$. Thus our model might allow a factor 2 gain in the
maximum wavenumber $k_{\rm max}$ up to which to include the power
spectrum information when constraining cosmological parameters. In the
sampling variance limited regime, the factor 2 gain is equivalent to a
factor 8 larger volume, yielding greater statistical power of the
power spectrum measurement. In addition, a wider coverage of
wavenumbers in the power spectrum amplitudes gives a higher
sensitivity to some of intriguing cosmological parameters such as the
total neutrino mass \citep{Saitoetal:08,Saitoetal:09,Saitoetal:11} and
the running index of the primordial power spectrum. Thus developing a
sufficiently accurate model of the nonlinear halo power spectrum is
very important in order for us to have improved cosmological
constraints, yet without having any significant biases in the derived
parameters.

Our model naturally predicts that, in the weakly nonlinear regime, the
halo power spectra show a scale-dependent bias relative to the
nonlinear matter power spectrum (see Figs.~\ref{fig:p_hm} and
\ref{fig:p_hh}).  The PT based model naturally predicts that the two
bias functions, defined as $b^{\rm cross}(k)=P_{{\rm h}m}(k)/P_{m}(k)$
and $b^{\rm auto}(k)=\sqrt{P_{{\rm hh}}(k)/P_{m}(k)}$, differ in the
weak nonlinear regime.  Furthermore, by incorporating the halo
occupation distribution (HOD) model, we can predict the nonlinear
power spectra of galaxy-matter and galaxy-galaxy in the weakly
nonlinear regime. We hope that our model can give a better description
of the nonlinear galaxy power spectra, and then allows for improved
cosmological constraints via the measured power spectra. Our model can
be further refined by including the higher-order loop corrections to
the nonlinear bias functions.

We showed that our model using the halo bias can give a good fit to
the simulation results for the halo spectra. This offers a promising
synergy between imaging and spectroscopic galaxy surveys, because a
cross-correlation of the spectroscopic galaxies with images of
background galaxies, the so-called galaxy-galaxy weak lensing, can
directly probe the mean mass of host halos and in turn constrain the
halo bias. Here the halo mass is constrained from the small-scale weak
lensing signals arising from the mass distribution within one halo,
which is complementary to the large-scale information of galaxy
clustering at $k\simlt 0.2~h{\rm Mpc}^{-1}$ we focus on in this
paper. This synergy is available if the spectroscopic and imaging
surveys see the same region of the sky. This is the case for upcoming
surveys: the BOSS and the Subaru HSC Survey, the Subaru HSC and PFS
surveys, the Euclid, the WFIRST and a combination of the LSST survey
with spectroscopic surveys.

However, our method rests on simplified assumptions one of which is
the local bias model.  Our model can be further improved by including
the non-locality of halo bias such as the dependence of halo bias on
the curvature of the initial density peaks \citep{Desjacquesetal:10}
and/or the tidal field around the density peaks \citep{Chanetal:12,
  Baldaufetal:12}. This is an interesting possibility, and will be
explored in our future work.

In this paper, we have focused on the real-space power spectra of
halos or galaxies. Actual observable for galaxy redshift survey is the
redshift-space power spectrum of galaxies, which is affected by the
redshift-space distortion effect due to peculiar motions of
galaxies. Towards a more accurate modeling of the nonlinear galaxy
power spectrum in redshift space, we need to further include the
nonlinear coupling between the redshift-space distortion effect and
the nonlinear galaxy bias.  There are encouraging developments in
modeling the redshift-space matter power spectrum in redshift-space,
based on the refined perturbation theory and $N$-body simulations
\citep[][]{Matsubara:08a,Taruyaetal:09,Taruyaetal:10,Tangetal:11,
  Matsubara:11,SatoMatsubara:11}. The redshift-space distortion effect
due to virial motions of galaxies within halos, the so-called
Fingers-of-God (FoG) effect, is harder to model, but
\citet{Hikageetal:12,Hikageetal:12b} recently developed an empirical
method to model the FoG effect based on the halo model and proposed a
method to remove the FoG contamination by combining with galaxy-galaxy
weak lensing measurement. It seems straightforward to incorporate
these methods in the method developed in this paper, in order to
include all the effects, nonlinear clustering, nonlinear bias,
nonlinear redshift-space distortion and FoG effect.  This is our
future work and will be presented elsewhere. These efforts are very
important in order to attain the full potential of future
high-precision galaxy surveys as well as to obtain unbiased, robust
cosmological constraints from the surveys.

%%%%%%%%%%%%%%%%%%%%%%%%%%%%%%%%%%%%%%%%%%%%%%%%%%%%%%%%%%%%%%%%%%%%%
%
\section*{acknowledgements}
%
%%%%%%%%%%%%%%%%%%%%%%%%%%%%%%%%%%%%%%%%%%%%%%%%%%%%%%%%%%%%%%%%%%%%%
We thank Issha Kayo, Ravi Sheth and Atsushi Taruya for useful
discussion.  In particular, we thank Atsushi Taruya for making the
code to compute the nonlinear matter power spectrum publicly available
to us.  This work is supported in part by the Grant-in-Aid for the
Scientific Research Fund (No. 23340061), by JSPS Core-to-Core Program
``International Research Network for Dark Energy'', by World Premier
International Research Center Initiative (WPI Initiative), MEXT,
Japan, and by the FIRST program ``Subaru Measurements of Images and
Redshifts (SuMIRe)'', CSTP, Japan. T. N. is supported by a
Grant-in-Aid for Japan Society for the Promotion of Science (JSPS)
Fellows (PD: 22-181).

\bibliographystyle{mn2e}
\bibliography{ms}

\end{document}